\documentclass[twocolumn,showpacs,amsmath,amssymb,showkeys,floatfix,prl]{revtex4}
\usepackage{amsfonts,amsmath}
\usepackage{graphicx}
\usepackage{dcolumn}
\usepackage{bm}

\begin{document}

\title{Physical region for three-neutrino mixing angles}

\author{D. C. Latimer and D. J. Ernst}

\affiliation{Department of Physics and Astronomy, Vanderbilt University, 
Nashville, Tennessee 37235, USA}
\date{\today}

\begin{abstract}

We derive a set of symmetry relations for the three-neutrino mixing angles, including the MSW matter effect. Though
interesting in their own right, these relations
are used to choose the physical region of the mixing angles such that
oscillations are parameterized completely and uniquely. 
We propose that the preferred way of setting the bounds on the mixing angles should be $\theta_{12} \in
[0,\pi/2]$, $\theta_{13} \in [-\pi/2,\pi/2]$, $\theta_{23}\in [0,\pi/2]$, and $\delta \in [0,\pi)$. No CP
violation then results simply from setting $\delta=0$.  In the presence of the MSW effect, this choice of bounds is a 
new result. Since the size of the asymmetry about $\theta_{13} = 0$ is dependent on the details of the data analysis and 
is a part of the results of the analysis, we argue that the negative values of $\theta_{13}$ should not be ignored. 

\end{abstract}



\maketitle
An abundance of data now demonstrates that neutrinos oscillate between flavor states.
Analyses of the data in the context of three-neutrino mixing can be found in 
Refs.~\cite{der,fog0,fog1,fog2,ahl1,wei,ahl2,ohl1,gon1,ahl3,gon3,fog4,fog3,gon2,lat1,gon4,mal,ble}.
Conventionally, the standard model of the electroweak 
interaction is extended to include a mass term for the neutrinos and a unitary mixing matrix 
which relates the flavor states of the neutrinos to the mass eigenstates.
We here utilize an algebraic formalism to derive symmetries of the mixing matrix which leave
the predicted oscillation probabilities invariant. In particular, we are interested in understanding
the bounds on the mixing angles which parameterize the mixing matrix. We derive the known bounds for the general case
which includes CP violation. For the case of no CP violation, we propose a new result, for the case in which the MSW effect is present,
of choosing the bounds as $\theta_{12} \in
[0,\pi/2]$, $\theta_{13} \in [-\pi/2,\pi/2]$, $\theta_{23}\in [0,\pi/2]$, $\delta = 0$. We argue that this
choice has distinct advantages for the phenomenological analysis of neutrino oscillation data.

The transformation from the mass states to the flavor states is represented by a unitary matrix $U_{\alpha j}$,
$\nu_\alpha = \sum_j U_{\alpha j} \nu_j$,
where we use Roman letters, $j$, for mass eigenstates and Greek letters, $\alpha$, for 
flavor states.
For two neutrinos, this mixing
matrix reduces to an element of the commutative group $U(1)$. For vacuum oscillations, the
mixing angles lie between 0 and $\pi/4$ in order to describe all discernible scenarios.  For propagation in matter,
this range on the mixing angle must be expanded to $[0,\pi/2]$.

In a three-neutrino theory, the mixing matrix is an element of a non-commutative group.
Symmetries in the parameterization are then less obvious, especially for neutrinos propagating in matter.
Transcendental expressions for locally defined mass eigenstates exist \cite{ohlsson4}, but
the underlying symmetries of these states are not transparent.
As such, we extend the algebraic formulation of vacuum neutrino oscillations developed in \cite{kchdcl1}
to include matter effects.
Thus we may derive symmetries among the mixing angles 
and the bounds on the mixing angles which these symmetries imply. The bounds on the mixing angles are well known in vacuum; 
for this case, the neutrino oscillation problem is then mathematically equivalent to the quark mixing problem. 

The limits on mixing angles, including the propagation through matter,
have previously been considered in \cite{gluza}.  Our 
results are in agreement with theirs;
however, our use of the symmetry relations permits us to arrive at additional
conclusions.  A discussion of the four neutrino 
case can be found in Ref.~\cite{pas}.

In \cite{kchdcl1}, the vacuum oscillation probability, valid for an arbitrary
fixed number of neutrino mass eigenstates, is given by
\begin{equation}
\mathcal{P}_{\alpha \to \beta}(t) =\frac{1}{2} \mathrm{tr}[P_+
e^{iHt}  P^\alpha
e^{-iHt} P_+ P^\beta ]\,\,,
\label{mixa}
\end{equation}
where $P^\alpha$ is the flavor projection operator given by $(P^\alpha)_{jk}
= (U^\dagger Q^\alpha U)_{jk}= U^*_{\alpha j} U_{\alpha k}$ 
in the mass-eigenstate basis, $P_+$ projects onto
the positive-energy states, and $H$ is an $n$-particle Dirac Hamiltonian for
masses $m_j$. We note that adding multiples of the identity to the
Hamiltonian leaves Eq.~(\ref{mixa}) invariant, and we recall that
the trace is cyclical.  
For the case of three neutrinos, the relativistic
limit, $p \gg m$, of the above equation is equivalent to the usual oscillation formula
\begin{eqnarray}
\mathcal{P}_{\alpha \to \beta}(L/E)&=& \mathrm{tr}
[U e^{i\mathcal{M}L/2E} U^\dagger Q^\alpha U e^{-i\mathcal{M}L/2E} U^\dagger Q^\beta]  
\nonumber \\ && \label{osctrace} \\ 
&=& \delta_{\alpha \beta}\nonumber\\
&&-4 \sum^3_{\genfrac{}{}{0pt}{}{j <
k}{j,k=1}} \mathrm{Re} (U_{\alpha j} U^*_{\alpha k} U_{\beta k} 
U^*_{\beta
j}) \sin^2 (\varphi_{jk})\nonumber \\
&&+2 \sum^3_{\genfrac{}{}{0pt}{}{j < k}{j,k=1}} \mathrm{Im} (U_{\alpha
j} U^*_{\alpha k} U_{\beta k}
U^*_{\beta j}) \sin(2 \varphi_{jk}) \,, \nonumber \\ &&\label{oscform}
\end{eqnarray}
where $\mathcal{M}$ is the diagonal matrix with entries $(m_1^2, m_2^2, m_3^2)$ and
$\varphi_{jk} := \Delta_{jk} L/4E$ with $\Delta_{jk} := m_j^2 -
m_k^2$. 

The generalization to neutrinos propagating in matter requires that the generator
of time (space) translations be modified to include a flavor-dependent potential
which accounts for interactions with the
electrons in matter \cite{msw}. The neutral current interaction does not effect oscillations and is neglected.
The charged current interaction involves only the electron flavor. 
As such, we must add to the Hamiltonian an effective potential 
which operates exclusively on the electron flavor.  Explicitly, this potential is
$A(x)=\sqrt{2} \,G \,E \,\rho(x)/m_n$, with $\rho(x)$ the electron density at position
$x$, $G$ the weak coupling constant, $m_n$ the nucleon mass, and $E$ the neutrino energy.  
In the mass basis, the dynamical equation in the presence of matter then becomes
\begin{equation}
i \partial_t \nu_m = (H + V) \nu_m = \widetilde{H} \nu_m, \label{mswdyn}
\end{equation}
where $V$ is given by $A(x) P^e$ with $P^e$ the electron flavor projection. 
Given nontrivial mixing among the mass and flavor states,
it is clear that the free Hamiltonian $H$ does not commute with the potential $V$.  

In the relativistic limit, one has $x \sim
t$ so that temporal derivatives are, in effect, spatial derivatives.
Formally, the solution to the differential equation in (\ref{mswdyn}) is
\begin{equation}
\nu_m(\Gamma) = \exp \left\{-i \int_\Gamma \widetilde{H} dx \right\} \nu_m(0), \label{mswexp}
\end{equation}
where $\Gamma$ parameterizes the oriented path of the neutrino through the matter with starting point $x=0$.  As the Hamiltonian is
hermitian, the adjoint of the above operator is simply $\exp \left\{ i \int_\Gamma \widetilde{H} dx \right\}$ which merely reverses
the orientation of the path $\Gamma$.

Written in a manner akin to that in (\ref{osctrace}), the expression for the oscillation probability in matter is
\begin{eqnarray}
\mathcal{P}_{\alpha \to \beta}(\Gamma,E) &=& \mathrm{tr} \left[
\exp \left\{ i \int_\Gamma U \widetilde{H} U^\dagger dx \right\}  Q^\alpha \right. \nonumber \\
&& \left. \exp \left\{ -i \int_\Gamma U \widetilde{H} U^\dagger dx \right\} Q^\beta \right]. \label{oscmass}
\end{eqnarray}
In the relativistic limit, the argument of the path ordered exponentials in this trace can be somewhat simplified
\begin{equation}
\int_\Gamma U \widetilde{H} U^\dagger dx = U\mathcal{M} U^\dagger L/2E + \int_\Gamma A(x) dx Q^e, \label{integral}
\end{equation}
modulo a multiple of
the identity; as noted above, multiples of the identity have no physical consequence with respect to oscillations.

We use the standard parametrization of the three-neutrino mixing matrix \cite{PDG}
\begin{widetext}
\begin{equation}
U(\theta_{23},\theta_{13},\theta_{12},\delta) = \left(  
\begin{array}{ccc}
c_{12} c_{13} & s_{12} c_{13} & s_{13} e^{-i \delta} \\
-s_{12}c_{23} - c_{12} s_{23} s_{13} e^{i \delta}  & 
c_{12} c_{23} - s_{12} s_{23} s_{13}  e^{i \delta}& s_{23}c_{13}\\
 s_{12}s_{23} - c_{12} c_{23} s_{13} e^{i \delta} & 
-c_{12} s_{23} - s_{12} c_{23} s_{13}  e^{i \delta}& c_{23}c_{13}
\end{array}
\right)\,\,, \label{mixer}
\end{equation}
\end{widetext}
where $c_{jk} = \cos{\theta_{jk}}$,
$s_{jk}=\sin{\theta_{jk}}$, and $\theta_{jk}$, $\delta$ are real.
We deem equivalent two quadruples of parameters $(\theta_{23}, \theta_{13}, \theta_{12}, \delta)$
if the oscillation probabilities
$P_{\alpha \to \beta}(\Gamma,E)$ are identical for all values of $\alpha$ and $\beta$ over some neutrino path $\Gamma$.  
This equivalence will be expressed via the relation $\equiv$.  Central to our method is the fact that the trace
of an operator is invariant
under conjugation by any unitary $U$,
\begin{equation}
\mathrm{tr} [U A U^\dagger ] = \mathrm{tr}[A]\,.\label{trprop}
\end{equation}
As the flavor projections $Q^{\alpha, \beta}$ in (\ref{oscmass}) must remain unchanged, they must commute 
with the conjugating matrix.

The group $SO(3)$ has three generators. It is convenient to 
express the mixing matrix in terms of the exponentiated generators $U_j(\theta)$, 
a rotation by angle $\theta$ about the $j$th
axis in $\mathbb{R}^3$.  To account for the CP phase, we make the additional definition 
for $S_\delta \in SU(3)$ by setting 
$S_\delta = \mathrm{diag} (e^{-i\delta/2},1,e^{i\delta/2})$ and letting
\begin{equation}
U_2^\delta(\theta)= S_\delta U_2(\theta) S_\delta^\dagger\,\,. \label{u2delta}
\end{equation}
The neutrino mixing matrix (\ref{mixer}) may be written as
\begin{equation}
U(\theta_{23},\theta_{13},\theta_{12},\delta)= 
U_1(\theta_{23}) U_2^\delta(\theta_{13}) U_3(\theta_{12})\,\,.
\end{equation}

From Eq.~(\ref{u2delta}), we may extract our first equivalence relation.  
For the CP phase $\delta=\pi$, one finds
\begin{equation}
S_\pi U_2(\theta) S_\pi^\dagger = U_2(-\theta)\,\,;
\end{equation}
hence, one has
\begin{equation}
U_2^{\delta+\pi}(\theta)=U_2^\delta(-\theta)\,\,.
\end{equation}
The addition of $\pi$ to the CP phase can accommodate a change in sign of $\theta_{13}$
\begin{equation}
(\theta_{23}, \theta_{13}, \theta_{12}, \delta) \equiv  (\theta_{23}, -\theta_{13}, 
\theta_{12}, \delta +\pi)\,\,. \label{deltapi}
\end{equation}
In order to proceed to further symmetries, we employ  
\begin{equation}
U_j(\pi)U_k(\theta)U_j(\pi)^\dagger = U_k(-\theta) \label{picommute}
\end{equation}
for $j \ne k$; this is easily confirmed by explicit calculation.  
Additionally, as $U_j(\pi)$ is diagonal, it commutes
with $S_\delta$ so that
\begin{equation}
U_j(\pi)U_2^\delta(\theta)U_j(\pi)^\dagger = U_2^\delta(-\theta) \label{picommute2}
\end{equation}
for $j=1,3$.  

For the same reason, $U_j(\pi)$ commutes with
the mass-squared matrix $\mathcal{M}$ and the projections $Q^\alpha$. 
The remaining step is to determine the action of $U_j(\pi)$ on the generator
of spatial translations.  Referring to Eq. (\ref{integral}), one sees that $U_j(\pi)$
commutes with the MSW potential and thus only affects the term containing the
mass-squared matrix $\mathcal{M}$.
Combining these facts with Eq.~(\ref{trprop}), we see that 
the mixing matrices $U$ and $U_j(\pi)^\dagger U U_j(\pi)$ yield identical oscillation probabilities.  
Using (\ref{picommute}) and
(\ref{picommute2}), we may express this equivalence in terms of the parameters themselves
\begin{subequations} 
\begin{eqnarray}
(\theta_{23}, \theta_{13}, \theta_{12}, \delta) &\equiv&
(-\theta_{23}, -\theta_{13}, \theta_{12}, \delta) \label{minus1}\\
&\equiv& (-\theta_{23}, \theta_{13}, -\theta_{12}, \delta) \label{minus2} \\ 
&\equiv& (\theta_{23}, -\theta_{13}, -\theta_{12}, \delta)\,\,. \label{minus3} 
\end{eqnarray}
\end{subequations}
From the same equations, we may determine the effect of adding $\pi$ to a mixing angle.  We find the following
equivalence amongst parameters
\begin{subequations}
\begin{eqnarray}
(\theta_{23}, \theta_{13}, \theta_{12}, \delta) &\equiv&
(\theta_{23} + \pi, \theta_{13}, \theta_{12},\delta) \label{pi1}\\ 
&\equiv& (-\theta_{23}, \theta_{13} + \pi, \theta_{12},\delta) \label{pi2}\\
&\equiv& (\theta_{23}, \theta_{13},\theta_{12} + \pi,\delta)\,\,. \label{pi3}
\end{eqnarray}
\end{subequations}

Trivially, the mixing angles satisfy a $2\pi$ periodicity; however, from
the relations in (\ref{pi1}--\ref{pi3}), it is clear that, without loss of generality, 
one may further restrict all $\theta_{jk}$ 
to lie \cite{ftn} in the interval $[0,\pi)$.  In the case of $\theta_{13}$, we are able to further
narrow these bounds to $[0,\pi/2]$.  An application of relation
(\ref{minus1}) followed by (\ref{pi2}) yields 
\begin{equation}
(\theta_{23}, \theta_{13}, \theta_{12}, \delta) \equiv
(\theta_{23} , \pi - \theta_{13}, \theta_{12},\delta)\,\,. \label{pihalf13}
\end{equation}
In general, given a real number $x$, the map \mbox{$x \mapsto a - x$} is a reflection 
about the point $a/2$ on the real line.  Hence, we see that if 
$\theta_{13}$ should lie between $\pi/2$ and $\pi$, then the above relation shows that
we have an equivalent oscillation probability for an angle reflected about $\pi/2$.  
In short, we may choose $\theta_{13}$ to lie in the first quadrant.  

For the remaining mixing angles, we may apply relations (\ref{minus2}), (\ref{pi1}), 
and (\ref{pi3}) to achieve
\begin{equation}
(\theta_{23}, \theta_{13}, \theta_{12}, \delta) \equiv
(\pi- \theta_{23} , \theta_{13}, \pi-\theta_{12},\delta)\,\,. \label{pihalf23}
\end{equation} 
From this, one notes that a reflection about $\pi/2$ in both $\theta_{23}$ 
and $\theta_{12}$ results in an equivalent oscillation probability.  These reflections
cannot be performed independently and still result in an equivalent theory.  Hence, we only 
have the freedom to restrict either $\theta_{23}$ or $\theta_{12}$ to the interval $[0,\pi/2]$, 
but not both.

Relation (\ref{pihalf23}) is valid for fixed $\delta$.  Should we relax this condition on $\delta$,
then we have the liberty to restrict both mixing angles to lie between $0$ and $\pi/2$.  
Using (\ref{deltapi}), (\ref{minus1}), and (\ref{pi1}), we have
\begin{equation}
(\theta_{23}, \theta_{13}, \theta_{12}, \delta) \equiv
(\pi- \theta_{23} , \theta_{13}, \theta_{12},\delta + \pi)\,\,.\label{lastm1}
\end{equation}
From similar logic, one can also deduce
\begin{equation}
(\theta_{23}, \theta_{13}, \theta_{12}, \delta) \equiv
(\theta_{23} , \theta_{13}, \pi - \theta_{12},\delta + \pi)\,\,.\label{last}
\end{equation}
Permitting a change in the CP phase results in the ability to independently reflect these two mixing
angles about $\pi/2$; hence, we are guaranteed, as noted in \cite{gluza, har}, that all $\theta_{jk}$ can be restricted to the
interval $[0,\pi/2]$ if one allows the full range on the CP phase $\delta \in [0, 2\pi)$.  
This is the common
understanding for the CKM mixing matrix for quarks and for neutrino oscillations including CP 
violation. 

We propose an alternate but equivalent set of bounds for the CP violating case. As a consequence of relation (\ref{deltapi}), 
we see that the full
range of the CP phase may be replaced by $\delta \in [0,\pi)$ by allowing a negative mixing angle $\theta_{13} 
\in [-\pi/2, \pi/2]$ 
with the two other mixing angles remaining in the first quadrant.   An alternative would be to utilize
Eq.~(\ref{pihalf23}) and allow either $\theta_{12} \in [0,\pi)$ {\it or} $\theta_{23} \in [0,\pi)$
with the CP phase in the first and second quadrants; however, given the smallness of $\theta_{13}$
in most models we find the extension to negative values of $\theta_{13}$ most natural.

Given that there is no indication of CP violation in neutrino oscillations, we give this case special
consideration. The conventional view of taking the the mixing angles $\theta_{jk} \in [0,\pi/2)$
requires that {\it both} $\delta = 0$ {\it and} $\delta = \pi$ must be included to cover the total allowed 
space for no CP violation, as was shown in \cite{gluza}. A common oversight has been to drop the
$\delta = \pi$ branch.  For no CP violation it is more
natural to have $\theta_{12} \in [0,\pi/2]$, 
$\theta_{13} \in [-\pi/2,\pi/2]$, $\theta_{23} \in [0,\pi/2]$, with only the $\delta = 0$ branch. This choice is motivated by the
smallness of $\theta_{13}$. We note that providing an upper limit on the value of $\sin^2 \theta_{13}$, a common practice \cite{PDG},
is insufficient since $\theta_{13} < 0$ is physical and the mixing is not symmetric in $\theta_{13}$. A two neutrino analysis
cannot distinguish the sign of $\theta_{13}$, but a linear term in $\theta_{13}$ is present for three neutrinos
and, depending on the details of the analysis, may be significant.

Experiments are not yet sensitive enough to determine the Dirac phase $\delta$. Until there exists data to
indicate a value of $\delta$ other than 0 or $\pi$ in the convention of Ref.~\cite{gluza}, or other than 0 
in the convention proposed here, the special case of no CP violation is the default assumption. An example of the
necessity, in principle, of incorporating the negative $\theta_{13}$ region arises from examining the question of mass hierarchy. 
To demonstrate this, we adopt the convention for
ordering the mass eigenstates by increasing mass, $m_1 < m_2 < m_3$.  This, together with the bounds chosen for
the mixing angles, defines a unique basis for an analysis, i.e. each possible physical solution is
included once and only once. Let us take as a typical set of parameters derived from
oscillation data,
$\Delta_{21} = 7 \times 10^{-5} \mathrm{eV}^2$, 
$\Delta_{32} = 3 \times 10^{-3} \mathrm{eV}^2$, 
$\theta_{12}=0.60, \quad \theta_{13}=0.10$, and
$\theta_{23}=0.80$.
This set of mass-squared differences is an example
of the regular hierarchy.  It is known that there is set of parameters with an inverted mass hierarchy that yield
oscillation probabilities which are nearly equivalent.  Using Eq.~(28) in \cite{flop}, we may calculate these
nearly equivalent parameters
$\Delta_{21} = 3 \times 10^{-3} \mathrm{eV}^2$,
$\Delta_{32} = 7 \times 10^{-5} \mathrm{eV}^2$,
$\theta_{12}= 1.45, \quad \theta_{13}= -0.60 $, and
$\theta_{23}= 0.70$.
We note the value of $\theta_{13}$ is negative and of relatively large magnitude.
This is a consequence of choosing a fixed (ascending) mass order.  Though we concede the common convention
for the inverted hierarchy is to reorder the masses,  
the point being made here is that the resolution of
the mass hierarchy question is independent of arbitrary mass ordering schemes. Utilizing
a single basis, as is usually done in quantum mechanics, requires negative values of $\theta_{13}$ (or the separate branch that arises from
$\delta = \pi$) to be mathematically complete and correct. 

Vacuum oscillations are a special case of oscillations in matter.  
Consequently, they automatically satisfy the above relations.  
The remaining question is to determine whether there exist any additional symmetries for the vacuum case,
 as is in the two-neutrino theory.  For three neutrinos, there are none.  In our above analysis, we note
that the operator $U_j(\pi)$ commutes with the MSW potential in Eq. (\ref{integral}).  As such, the presence of this
potential is inconsequential in the derivation of the symmetries that follow; hence, the special case of vanishing
electron density admits no additional symmetries. To ensure that we have not overlooked a symmetry which could further reduce the bounds
on the mixing angles, we have checked numerically that the mixing probabilities are indeed 
unique over the regions given.

In summary, we have put forth a new method for deriving symmetries of the three-neutrino mixing matrix.
These symmetries are of interest themselves. They also provide an elegant way to determine the
physical bounds on the mixing angles in the presence of the MSW effect.  In particular, we note that a convenient way of setting these bounds
is to take $\theta_{13}\in [-\pi/2,\pi/2]$, $\delta \in [0,\pi)$, and the other two angles in the first quadrant. In the case of no CP violation,
where some amount of confusion appears in the literature, this reduces to setting $\delta=0$, and taking 
$\theta_{13}\in [-\pi/2,\pi/2]$ with the other two angles in the first quadrant. We note that in an expansion in terms of the ratio of the small mass square difference to
the large mass square difference, the lowest term is dependent only on $\sin^2\theta_{13}$. In this limit, negative $\theta_{13}$ would be redundant.
However, the leading correction is linear in $\theta_{13}$. We examine this question in Ref.~\cite{nolsnd} and find that, for reasonable parameters, the
$\chi^2$ space may have a noticeable asymmetry about $\theta_{13}=0$, thus further motivating that the parameter space for the mixing angles
be taken as derived here, a result that in the presence of the MSW effect, has not previously appeared in the literature.

\bibliography{bothang}

\begin{thebibliography}{1}
\bibitem{ftn} We utilize the curved bracket $)$ to indicate that the upper limit
is not included within the range and utilize the square bracket ] when the limit 
is a part of the range.
\end{thebibliography}

\end{document}